\documentclass[epj]{svjour}

\def\Za{Z\alpha}
\def\Zab{(Z\alpha)}
\def\En{{\cal E}_{2s}}
\def\eps{\epsilon}

\def\kp{\kappa}
\def\kpt{\tilde{\kappa}}

\catcode`@=11  

\def\@calc#1=#2#3#4{
#1=#2
\if#3+\advance#1 by #4\fi
\if#3-\advance#1 by -#4\fi
\if#3*\multiply#1 by #4\fi
\if#3/\divide#1 by #4\fi
}
\newcount\@a  
\newcount\@b  %
\newcount\@c  %
\newcount\@d  %

\newcommand{\dgap}{5} 
\newcommand{\ov}{10} 
\newcommand{\ovv}{20} 
\newcommand{\ovvv}{30} 
\newcommand{\ovvvv}{40} 
\newcommand\dashgap{20}  
\newcommand\dashlen{10}  

\newcommand{\Cross}{
   \put(0,0){\kern-0.39em\lower0.6ex\hbox{$\times$}}
}

\newcommand{\SquareH}{
\kern -.5em \lower .25em \hbox{\vrule width .5em height .5em}
\Large \kern -1em \lower 1.5em \hbox{$H$}
}

\newcommand{\CrossT}{
\Cross
\Large \kern -.3em \lower 1.5em \hbox{$T$}
}

\def\Hline(#1,#2,#3){ 
 \put(#1,#2){\line(1,0){#3}}
}

\def\dHline(#1,#2,#3){ 
 \multiput(0,0)(0,\dgap){2}{\put(#1,#2){\line(1,0){#3}}}
}

\def\thickHline(#1,#2,#3){ 
 \put(#1,#2){\vrule width #3\unitlength height 10\unitlength}
}

\def\Vwaveline(#1,#2,#3){ 
 \@calc\@a={#3}/\ovvvv
 \multiput(#1,#2)(0,\ovvvv){\@a}{
    \put(0,\ov){\oval(\ovv,\ovv)[l]}
    \put(0,\ovvv){\oval(\ovv,\ovv)[r]}
 }
}

\def\Vdashline(#1,#2,#3){ 
 \@calc\@c={#3}/\dashgap
 \multiput(#1,#2)(0,\dashgap){\@c}{\line(0,1){\dashlen}}
}

\catcode`@=12 

\begin{document}
\title{Hyperfine Structure of the Ground and First Excited States\\ in Light Hydrogen-Like Atoms and
High-Precision Tests of QED}
\author{Savely G. Karshenboim\inst{1,2}\thanks{E-mail: sek@mpq.mpg.de} \and Vladimir G. Ivanov\inst{3,1}
}                     
\institute{D. I. Mendeleev Institute for Metrology, 198005 St. Petersburg, Russia
\and Max-Planck-Institut f\"ur Quantenoptik, 85748 Garching, Germany
\and Pulkovo Observatory, 196140, St. Petersburg, Russia
}
\date{Received: date / Revised version: date}
\titlerunning{HFS in Light Hydrogen-Like Atoms}
\authorrunning{S. G. Karshenboim and V. G. Ivanov}
%
\abstract{
We consider hyperfine splitting of $1s$ and, in part, of $2s$ levels in light
hydrogen-like atoms: hydrogen, deuterium, tritium, helium-3 ion, muonium
and positronium. We discuss present status of precision theory and
experiment for the {\em hfs} intervals. We pay a special attention to a
specific difference, $D_{21}  = 8 E_{\rm hfs}(2s) - E_{\rm hfs}(1s)$, which
is known experimentally for hydrogen, deuterium and $^3{\rm He}^+$ ion.
The difference is weakly affected by the effects of the nuclear structure and thus
may be calculated with a high accuracy. We complete a calculation of the fourth order QED contributions to this
difference and present here new results on corrections due to the nuclear effects. Our theoretical predictions
appear to be in a fair agreement with available experimental data.
Comparison of the experimental data with our examination of $D_{21}$
allows to test the state-dependent sector of theory of the {\em hfs} separation of the $1s$ and $2s$
levels in the light hydrogen-like atoms up to $10^{-8}$.
} 
\maketitle

\section{Introduction}

The hyperfine structure ({\em hfs}) interval in the ground state of a
number of simple atoms (hydrogen \cite{17h1s}, deuterium \cite{17d1s},
tritium \cite{mathur} and helium-3 ion \cite{17he1s}) has been measured
with a high precision. The {\em hfs} separation in the $2s$ metastable state
in hydrogen \cite{17h2s,17rothery}, deuterium \cite{17d2s} and the $^3$He$^+$
ion \cite{17he2s,17prior} was also measured accurately.
Some experimental results are as old as almost fifty
years, but the accuracy of even present-day theoretical calculations for the
{\em hfs} interval in those light atoms is much lower than that
for the experiments (see e.g. Table~\ref{T1shfs}). Theory of the {\em hfs} interval in simple atoms is
essentially based on the bound state Quantum Electrodynamics ({\em QED}),
however effects due to the nuclear structure are unavoidable and they
strongly affect the energy levels. Their uncertainty limits the theoretical accuracy for
the hyperfine splitting in hydrogen, deuterium, tritium and helium-3 ion
on a level of 10-200 ppm.

One of ways to avoid the problem of the nuclear effects is to study atoms
free of any nuclear structure such as muonium and positronium. Hyperfine
splitting in these pure leptonic atoms was measured
\cite{MuExp,MuExp1,PsExp1,PsExp2} with an accuracy appropriate for
precision tests of the bound state QED. Other possibilities to avoid the
problem of lack of accurate knowledge of the corrections induced by the
nuclear effects is related to a fact that those corrections are
proportional to the squared value of the wave function at the origin
\begin{eqnarray}
  \Delta E({\rm Nucl}) &=& A({\rm Nucl})\times |\Psi_{nl}({\bf r}=0)|^2\;,\label{NuclPsi1}\\
  \Psi_{nl}({\bf r}=0) &=& \frac{(Z\alpha)^3 m_R^3}{\pi n^3}\delta_{0l}\label{NuclPsi2}\;,
\end{eqnarray}
where $\alpha$ is the fine structure constant, $Z$ is the nuclear charge
and $m_R$ is the reduced mass of the orbiting particle. The relativistic
units in which $\hbar=c=1$ are used here and through the paper. We ignore
a difference between energy interval $E$ and a measured frequency $\nu=E/h$:
presenting the theoretical expression for the energy splitting $E$ and numerical
results for the frequency $\nu$.
Here $\Psi_{nl}({\bf r})$ is the Schr\"odinger-Coulomb wave function and
$A({\rm Nucl})$ is a nuclear parameter which does not depend on the atomic
state $nl$. Comparing the {\em hfs} for the atoms with a different value
of $\Psi_{nl}(0)$ one can reduce influence of the nuclear structure and
test the bound state QED with a high accuracy. There are two options to
vary $\Psi_{nl}(0)$:
\begin{itemize}
\item
to compare muonic and electronic atoms (i.e. to study atoms with the same
nucleus and different values of $m_R$);
\item
to compare {\em hfs} intervals for the $ns$ states with different value of
$n$ or to study the hyperfine splitting for states $l\neq 0$.
\end{itemize}
Presently accurate experimental data are available only for one of these two
options: it is possible to take advantage of existence of precision experimental
data on the $1s$ and $2s$ hyperfine intervals in a few of light two-body atomic systems.
A comparison of the data for the $1s$ and $2s$ {\em hfs} intervals allows to determine value of
a specific difference
\begin{equation}
  D_{21}  = 8 E_{\rm hfs}(2s) - E_{\rm hfs}(1s)
\end{equation}
in hydrogen, deuterium and helium-3 ion. The theory of this specific
difference can be developed much more successfully than that for the ground state
interval $E_{\rm hfs}(1s)$ because of the essential cancellation of the
nuclear effects (see Eqs.~(\ref{NuclPsi1}, \ref{NuclPsi2}).

The most accurate experimental value for $D_{21}$ was obtained for the helium-3 ion
\begin{equation}\label{d21he3exp}
D_{21}^{\rm exp}(^3{\rm He}^+) = 1\,189.979(71)~{\rm kHz}
\end{equation}
after comparison of results obtained for the $1s$ state in 1969 \cite{17he1s} and for the $2s$ state in 1977
\cite{17prior}. The QED theory was developed by that time up to third
order corrections including the $(Z\alpha)^2E_F$, $\alpha(Z\alpha)^2E_F$
and $(Z\alpha)^2(m/M)E_F$ contributions (here $E_F$ is the so-called Fermi
energy, leading contribution to the $1s$ {\em hfs} separation).
The experimental result in Eq.~(\ref{d21he3exp}) happened to be in some agreement
with theory, however, uncertainty of theory was not properly estimated. 
Here we present new theoretical results on $D_{21}$ in hydrogen, deuterium and helium-3 ion 
\cite{newd21}.
In our paper we demonstrate that there are a number of higher-order
QED corrections which were not taken into account and which are competitive
with the uncertainty of the experiment. We complete calculation of fourth
order corrections and present theoretical results with accuracy higher
than that for the measurements. The higher-order nuclear-structure effects
also contribute to the difference $D_{21}$ and their contribution is
important for a comparison with experiment. They are considered in our paper
in detail.


\begin{table*}[bt]
\caption{The ground state {\em hfs} interval in hydrogen, deuterium,
tritium and helium-3 ion.}
\label{T1shfs}
\begin{center}
\def\arraystretch{1.4}
\setlength\tabcolsep{5pt}
\begin{tabular}{lcccc}
\hline
Atom & $E_{\rm hfs}^{\rm exp}$& $E_{\rm hfs}^{\rm QED}$& $E_{\rm hfs}^{\rm exp}-E_{\rm hfs}^{\rm QED}$
& $(E_{\rm hfs}^{\rm exp}-E_{\rm hfs}^{\rm QED})/E_F$\\
&[kHz]&[kHz]&[kHz]&[ppm]\\
\hline
Hydrogen & ~~~1\,420\,405.751\,768(1), \protect\cite{cjp2000}& ~~1\,420\,452& ~-46&-33\\
Deuterium & ~~~~327\,384.352\,522(2), \protect\cite{17d1s}~& ~~~~327\,339&~~45 &138\\
Tritium & ~~1\,516\,701.470\,773(8), \protect\cite{mathur}~& ~~1\,516\,760& ~-58&-38\\
$^3$He$^+$ ion & - 8\,665\,649.867(10),~~~~ \protect\cite{17he1s}~& -8\,667\,569& 1919&221\\
\hline
\end{tabular}
\end{center}
\end{table*}

The paper is organized as following: in Sect. 2 we consider the QED theory of $1s$ {\em hfs} interval
and determine parameters $A({\rm Nucl})$ for hydrogen, deuterium and
$^3{\rm He}^+$ ion. Sect. 3 is devoted to QED calculations of the difference
$D_{21}$. We study the fourth order QED contributions and, in particular, we
find the vacuum polarization contribution in order $\alpha(Z\alpha)^3E_F$ and
the leading logarithmic recoil term in order $(Z\alpha)^3(m/M)E_F$.
The nuclear effects are taken into account in Sect. 4. We show that study
of the difference $D_{21}$ provides an effective test of QED theory of the
{\em hfs} intervals $E_{\rm hfs}(1s)$ and $E_{\rm hfs}(2s)$ on a level of
accuracy essentially below 1 ppm and such a test is free of problems of
the nuclear structure. That is quite competitive with investigations of
the hyperfine splitting in the ground state of muonium and positronium and we present
a brief overview of them in Sect. 5. Section 6 summarized the paper and
a comparison of theory
and experiment is presented there for the difference $D_{21}$ in hydrogen, deuterium and
helium-3 ion and for the ground state {\em hfs} separation in muonium and positronium.

\section{Hyperfine splitting in the ground state\\ in hydrogen, deuterium and helium ion}

The hyperfine splitting of an $ns$ state in a hydrogen-like atoms is determined
in the non-relativistic approximation by the so-called Fermi energy:
\begin{eqnarray}
  E_{\rm hfs}(ns) &=& {E_F\over n^3}\;,\\
  E_F/h &=&
  {8 \over 3} \,Z^3 \alpha^2 \,c\cdot Ry \, {\mu \over \mu_B}
  \,{ 2 I + 1 \over 2 I}\,
\left({ M\over m+M}\right)^3\label{fermiE}\;.
\end{eqnarray}
Here $Ry$ is Rydberg constant, $c$ is speed of light, $h$ is the Planck
constant, $\mu_B$ is the Bohr magneton, $m$ is the electron mass, $M$ is
the nuclear mass and $I$ is the nuclear spin. The nuclear magnetic moment
$\mu$ in our notation can be negative (if its direction is opposite to the
nuclear spin) and the Fermi energy $E_F$ defined as the splitting between
states with atomic angular moments $F=I+1/2$ and $I-1/2$, calculated within the
non-relativistic approximation, can be negative as well.

The result of the QED calculations is of the form
\begin{equation}\label{QED1sF}
  E_{\rm hfs}^{\rm QED}(1s) = E_F \bigl(1+Q_{QED}(1s)\bigr)\;,
\end{equation}
and
\begin{eqnarray}\label{QED1s}
Q_{\rm QED}(1s)&=&a_e+\left\{\frac{3}{2}(Z\alpha)^2+ \alpha(Z\alpha)\left(\ln2-\frac{5}{2}\right)\right.\nonumber\\
&+&{\alpha (Z \alpha )^2\over \pi}\left[-\frac{2}{3}\ln{1\over(Z\alpha)^2}
\left(\ln{\frac{1}{(Z\alpha)^2}}\right.\right.\nonumber \\
&+&\left.4\ln2-\frac{281}{240}\right) +17.122\,339\ldots \nonumber \\
&-&\left.\left.\frac{8}{15}\ln{2}+\frac{34}{225}\right]+0.7718(4)\,\frac{\alpha^2(Z\alpha)}{\pi}\right\}\,.
\end{eqnarray}
The references to all terms can be found in a review in Ref.~\cite{EGS}. The expression above is a result of
the external field approximation. The recoil corrections involve
integration over high momentum $k\sim M$ and a consideration of a nucleus
as a point-like one is not valid in such a case.
Actually a theory of a point-like
particle with an anomalous magnetic moment is inconsistent and leads to
a divergency at high momentum transfer for the nuclear vertex. The muon (and
electron) must be treated as either a point-like particle without anomalous moment,
or a particle with the anomalous magnetic moment and an internal structure.
The magnetic anomaly and ``structure'' effects come from the same diagrams. As a result
one need to be very careful when considering a ``point-like'' nucleus as an approximation.
It is important to mention that the leading recoil
corrections \cite{17arno} and most of non-leading terms involve the
nuclear structure effects and proportional to the $|\Psi_{nl}(0)|^2$ (cf. (\ref{NuclPsi1})).

To compute any numerical result we use here $\alpha^{-1}=137.036\,000$, $c\cdot Ry=3.289\,841\,960\cdot
10^{12}$~kHz, $a_e=1.159\,652\cdot10^{-3}$ and
parameters of nuclei collected in Table~\ref{Tpara}. The values of the
fundamental constants and nuclear parameters are based on data taken from
\cite{17mohr,firestone}, but we keep them here only with the accuracy
sufficient for our purposes. The results of calculation are summarized in
Table~\ref{T1shfs}. Leading recoil corrections depend on nuclear
structure. Some pure QED corrections of higher order are known but not
included being essentially smaller than uncertainty of nuclear effects.


\begin{table*}[bt]
\caption{Parameters for calculations of the {\em hfs} interval in hydrogen, deuterium and helium-3 ion
\protect\cite{17mohr,firestone}. The proton charge radius is taken from Ref.~\protect\cite{17kars99}.}
\label{Tpara}
\begin{center}
\def\arraystretch{1.4}
\setlength\tabcolsep{5pt}
\begin{tabular}{lccccccc}
\hline
Atom       & $Z$& $I$& $M/m$  & $\mu/\mu_B$ & $E_F$ & $\eta$&$R_E$  \\
  & & & &  [$10^{-3}$]& [kHz]& &[fm] \\
\hline
Hydrogen   & 1& 1/2 &1\,836.153 & ~1.521\,032\,2 &~1\,418\,840 & 5.585\,69&0.88(3) \\
Deuterium  & 1& 1   &3\,670.483 & ~0.466\,975\,5 & ~~~326\,968 &  1.714\,03& 2.13(1) \\
Tritium   & 1& 1/2 &5\,496.922 & ~1.622\,393\,6 & ~1\,515\,038 &  17.831& \\
$^3$He$^+$ & 2& 1/2 &5\,495.885 &  -1.158\,750\,5 & -8\,656\,598&  -6.368\,36&1.67(1) \\
\hline
\end{tabular}
\end{center}
\end{table*}

The theoretical calculations above take into account only pure QED terms, while the nuclear
effects can be estimated via a comparison of the experiment and the pure QED theory
\begin{eqnarray}
  E^{\rm Nucl}_{\rm hfs}(1s)&=&E^{\rm Exp}_{\rm hfs}(1s)-E^{\rm QED}_{\rm hfs}(1s)\;,\label{Nucl1s1}\\
  A_{\rm hfs}({\rm Nucl})&=&\frac{E^{\rm Exp}(1s)-E^{\rm QED}(1s)}{|\Psi_{1s}(0)^2|}\label{Nucl1s2}\;.
\end{eqnarray}
The nuclear models or study experimental data on nuclei offer another way to
find $E^{\rm Nucl}(1s)$ and $A({\rm Nucl})$ and they are discussed in part in Sect.~\ref{SectNucl}.

\section{QED calculations of $D_{21}$ in light atoms}

The evaluation of the QED corrections involves contributions of the second,
third and fourth order in unit of the Fermi energy. The  second
\cite{17brei} and third \cite{17zwanziger,17sternheim,17pmohr} order
corrections were calculated some time ago (see Table~\ref{17Tab13}). One of the
fourth order corrections, $(Z\alpha)^4E_F$, was also found that time \cite{17brei}.
Other fourth order terms were found only recently and the theoretical expression is
now of the form
\begin{eqnarray}\label{d21mod}
D_{21}({\rm QED}) &=& (Z\alpha)^2 \,E_F
\times \Biggl\{ \left[{5 \over 8} + {177 \over 128}\,(Z\alpha)^2\right]
\nonumber\\
&+& {\alpha\over \pi} \, \left[
\left( {16 \over 3} \,\ln2-7\right)\, \ln(Z\alpha) -5.221\,23\dots\right]
\nonumber\\
&+& {\alpha\over \pi}\,\left[{8 \over 15} \,\ln2 - {7 \over 10}\right]
\nonumber\\
&+& {m\over M}\,\biggl[
 -{9 \over 8} +  \left({ \ln2 \over 2}-{7 \over 32}\right)\left(1-{1\over \eta}
 \right)\nonumber\\
&~&\qquad -\left({145 \over 128} - { 7 \over 8} \,\ln2 \right)
\eta
\biggr]\nonumber\\
&+&{\alpha^2\over2\pi^2}\,
\left({16\over3}\,\ln2-7\right)\,\ln(Z\alpha)\nonumber\\
&-&{\alpha\over\pi}{2m\over M}\,\left({16\over3}\,\ln2-7\right)\,\ln(Z\alpha)\nonumber\\
&+&\frac{Z\alpha}{\pi}{m\over M}\left({4\over 3}\ln2-2\right)\ln(Z\alpha)\nonumber\\
&+&\alpha(Z\alpha)\Bigl(C_{\rm SE}+C_{\rm VP}\Bigr)\Bigg\},
\end{eqnarray}
where
\begin{equation}
  \eta = {\mu \over \mu_B} \,{M \over m}\,{1 \over {Z\, I}}\;.
\end{equation}
Two corrections in the fourth order, $\alpha(Z\alpha)^2(m/M)E_F$
and $\alpha^2(Z\alpha)^2E_F$, were found in Ref.~\cite{preliminary}
in the leading logarithmic approximation and their uncertainties
are estimated by a half-value of the leading logarithmic terms.
The coefficients $C_{\rm SE}$ and $C_{\rm VP}$ related to
the self-energy and vacuum polarization higher-order radiative corrections
were first estimated in Ref.~\cite{preliminary}, but with some misprints.
Below we correct that estimation and discuss a recent calculation in
Ref.~\cite{yero2001}.

The expression takes onto account some recoil effects. As
it was demonstrated by Sternheim \cite{17sternheim} the $n$-dependent part
of the $(Z\alpha)^2(m/M)E_F$ contribution into $E_{\rm hfs}(ns)$ does not
depend on the nuclear structure. In the case of the $\alpha(Z \alpha)^2(m/M)E_F$ and
$(Z \alpha)^3(m/M)E_F$ that is correct at least for the logarithmic terms. The pure
recoil logarithmic correction $(Z \alpha)^3(m/M)\ln(Z \alpha)E_F$
is evaluated here.
A logarithmic part of the QED correction in order $(Z\alpha)^3m/M$ is
easy to calculate with help of effective potentials which are responsible for $(Z\alpha)^5m^2/M$
correction to the Lamb shift (cf. \cite{17Zh,17ZP}). The result is
\begin{equation}\label{leadrec}
 2\cdot\frac{2}{3}
 \frac{(Z\alpha)^5}{\pi}
 \frac{m^3}{M}
 \ln\left(\frac{1}{Z\alpha}\right)
 \frac{E_F}{(Z\alpha m)^2}\cdot
 \left(\frac{3}{2}-\ln2\right)
 \;.
\end{equation}
However this result is of a reduced value as far as the effective potential for
the Lamb shift in order $(Z\alpha)^5m^2/M$ has been used. The logarithmic
term (with $\ln(Z\alpha)$) in that order is not dominant. That often happens with pure recoil
contributions (in contrast the logarithmic terms dominates in the case of
most of radiative and radiative-recoil corrections).
The other logarithmic contribution in order $(Z\alpha)^5m^2/M$ (with a recoil logarithm $\ln(mR)$) and a part of
the non-logarithmic term
are effectively included into the nuclear-structure contributions (see the next section).
An essential non-logarithmic part which is not
included there is related to two-photon effective potentials with derivatives.
The scale of the loop integration momentum is determined by the electron
mass and the related contribution does not depend on the nuclear
structure (cf. the Sternheim contribution into Eq.~(\ref{d21mod}) in order $(Z\alpha)^2(m/M)E_F$).
It can be essentially enhanced because of a big value of the nuclear anomalous magnetic moment and we
estimate that non-logarithmic contribution as $\pm\eta(Z\alpha)^3/\pi\,m/M \,E_F$.


%
\begin{table}[bth]
\caption{QED contributions up to third order to the $D_{21}$ in hydrogen, deuterium and
helium-3 ion.}
\label{17Tab13}
\begin{center}
\def\arraystretch{1.4}
\setlength\tabcolsep{5pt}
\begin{tabular}{lccc}
\hline
Contribution                        &  H      & D        & $^3$He$^+$  \\
&[kHz]&[kHz]&[kHz]\\
\hline
$(Z\alpha)^2 E_F$        & 47.222\,0 &  10.882\,2 & -1\,152.439\,0 \\
$\alpha(Z\alpha)^2 E_F $ (SE)      &  ~1.936\,0 &   ~0.446\,1 & ~~~-37.441\,5     \\
$\alpha(Z\alpha)^2 E_F $ (VP)      & -0.058\,0 &  -0.013\,4 & ~~~~~1.414\,8     \\
$(Z\alpha)^2{m\over M} E_F $       & -0.162\,9 &  -0.009\,4 & ~~~~-0.796\,7      \\
\hline
Total up to 3rd order & 48.937\,1 & 11.305\,6 & -1\,189.262\,4\\
\hline
\end{tabular}
\end{center}
\end{table}

\begin{figure}[bth]
\begin{center}
\noindent
{
\begin{picture}(360,450)
 \Hline(0,240,390)
 \Vwaveline(180,-6,80)
 \put(180,-6){\Cross}
 \put(180,117){\circle{80}}
 \Vwaveline(180,160,80)
 \put(190,-100){\kern-0.5em {\em a}}
\end{picture}
}\\
\end{center}
\begin{center}
{
\begin{picture}(360,450)
 \Hline(0,240,390)
 \thickHline(60,235,250)
 \Vdashline(60,0,80)
 \put(60,0){\Cross}
 \put(60,115){\circle{80}}
 \Vdashline(60,160,80)
 \Vwaveline(300,0,240)
 \put(300,0){\Cross}
\end{picture}
}
\begin{picture}(70,450)
\put(20,100){\Large+}
\put(50,-100){\kern-0.5em {\em b}}
\end{picture}
{
\begin{picture}(360,450)
 \Hline(0,240,390)
 \thickHline(68,235,234)
 \Vwaveline(60,0,240)
 \put(60,0){\Cross}
 \Vdashline(300,0,80)
 \put(300,0){\Cross}
 \put(300,115){\circle{80}}
 \Vdashline(300,160,80)
\end{picture}
}
\vskip .7cm
\end{center}
\caption{\label{fig1}Vacuum polarization contributions to {\em hfs}}
\end{figure}
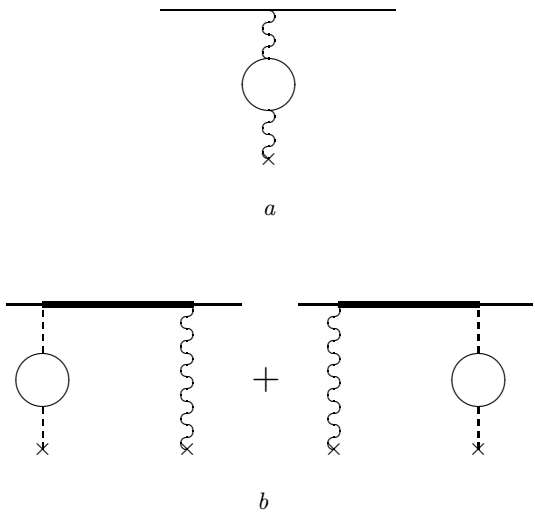

The higher-order vacuum polarization correction related to $C_{VP}$ is found in this paper. The
contribution comes from diagrams depicted in Fig.~1. They are evaluated with
the exact Dirac wave functions and the Green function of an electron in the Coulomb
field and the result is expanded in powers of $Z\alpha$. The results for
the vacuum polarization contribution to the $2s$ {\em hfs} interval are (cf.
Ref.~\cite{jetp00,17kis}):
\begin{eqnarray}
 \Delta E({\rm Fig.~1a})&=&\frac{\alpha}{\pi}\frac{E_F}{8}\times
 \Biggl[\frac{3\pi}{8}\,\Za-\frac{7}{10}\,\Zab^2\nonumber\\
 &+&\Zab^3\left(\frac{143\pi}{192}-\frac{3\pi}{8}
 \,\ln\left(\frac{\Za}{4}\right)\right)\nonumber\\
 &~&\qquad +\dots\Biggr]\,,\nonumber\\
 \Delta E({\rm Fig.~1b})&=&\frac{\alpha}{\pi}\frac{E_F}{8}\times
 \Biggl[\frac{3\pi}{8}\,\Za\nonumber\\
 &+&\Zab^2\left(\frac{34}{225}-\frac{8}{15}
 \,\ln(\Za)\right)\nonumber\\
 &+&\Zab^3\left(\frac{1715\pi}{1152}-\frac{\pi}{6}
 \,\ln\left(\frac{\Za}{4}\right)\right)\nonumber\\
 &~&\qquad +\dots\Biggr]\nonumber\,.
\end{eqnarray}
Finally one can obtain
\begin{eqnarray}\label{eFig1}
\Delta E^{\rm VP}_{\rm hfs}(2s)&=&\frac{\alpha}{\pi}\frac{E_F}{8}
\times\Biggl[\frac{3\pi}{4}\,\Za\nonumber\\
&+&\Zab^2\left(-\frac{247}{450}-\frac{8}{15}\,\ln(\Za)\right)\nonumber\\
&+&\Zab^3\left(\frac{2573\pi}{1152}-\frac{13\pi}{24}
\,\ln\left(\frac{\Za}{4}\right)\right)\nonumber\\
&~&\qquad +\dots\Biggr]\,.
\end{eqnarray}
The derivation is considered in detail in Appendix~\ref{appendixVP}. To find a correction
to the difference $D_{21}$ one has to compare the result for the vacuum polarization
contribution to the $2s$ {\em hfs} obtained above with that for the ground state 
\cite{jetp00,17kis}
\begin{eqnarray}
\Delta E^{\rm VP}_{\rm hfs}(1s)&=&\frac{\alpha}{\pi}\,E_F\times\Biggl[\frac{3\pi}{4}\,\Za\nonumber\\
&+&\Zab^2\left(\frac{34}{225}-\frac{8}{15}\,\ln(2\Za)\right)\nonumber\\
&+&\Zab^3\left(\frac{539\pi}{288}-\frac{13\pi}{24}\,\ln\left(\frac{\Za}{2}\right)\right)\nonumber\\
&~&\qquad+\dots\Biggr]\;.
\end{eqnarray}
Finally we find
\begin{eqnarray}\label{d21vp}
 \Delta D_{21}^{\rm VP}=\frac{\alpha}{\pi}E_F\times
 \Biggl\{\Zab^2\left(-{\frac{7}{10}}+{\frac{8}{15}}\,\ln(2)\right)\nonumber\\
 +\Zab^3\left(\frac{139\pi}{384}+\frac{13\pi}{24}\,\ln(2)\right)+\dots\Biggr\}
\end{eqnarray}
and thus
\begin{equation}\label{numcvp}
  C_{\rm VP}= \frac{139}{384} + \frac{13}{24} \, \ln{2}\simeq 0.74\;.
\end{equation}
As we have mentioned some partial
results on $\alpha(Z\alpha)^3E_F$ terms were presented in Ref.~\cite{preliminary}
with a misprint. The relative sign in Eqs. (24,25,27) of paper \cite{preliminary}
is to be corrected and the result is (the corrected sign is marked by $*$)
\begin{eqnarray}\label{correczed}
C_{\rm SE}&=&\left[{139\over 16 }-4\ln{2}\right]\left[ {3\over 2}-\ln{2}\right]\nonumber\\
&~&\qquad
-^{\mbox{\bf *}}\left[ {13\over 4} -\ln{2}\right]\left[ \ln(2)+{3\over 16}\right]
\;,\nonumber\\
C_{\rm VP}&=&{5\over 24}\left[{3\over 2}-\ln{2} \right]+^{\mbox{\bf *}}{3\over 4}
\left[ \ln{2}+{3\over 16}\right]\nonumber\;.
\end{eqnarray}
It was also then expected \cite{preliminary} that the
$\alpha(Z\alpha)^3E_F$ results found there are likely incomplete. Recently
Yerokhin and Shabaev directly calculated the self-energy
contribution \cite{yero2001} after our suggestion
\begin{eqnarray}\label{numcse}
 C_{\rm SE}(Z=1)&=&2.07(25)\;,\nonumber\\
 C_{\rm SE}(Z=2)&=&2.01(19)\;.\nonumber
\end{eqnarray}
The self-energy result of Ref.~\cite{yero2001} is affected by the higher order corrections and
thus slightly depends on $Z$.

The complete results (\ref{numcvp}) and (\ref{numcse}) indeed disagree
with the corrected above partial results in Eqs.~(\ref{correczed})
\begin{eqnarray}\label{numpart}
 C_{\rm SE}&=&{795\over 64}-{7\over4}+5\ln^2{2}\simeq 2.5\;,\nonumber\\
 C_{\rm VP}&=&{29\over 64}+{13\over24}\ln{2}\simeq 0.83\;.
\end{eqnarray}
and in part it is caused by appearance of effective
nonrelativistic operators with derivatives. Those operators
do not contribute into
logarithmic corrections to ground state hyperfine structure
\cite{17ZP,17Zh} and were not considered in Ref.~\cite{preliminary}. The
difference for complete and partial results
is numerically small for both: the vacuum polarization and self energy. That is related
to the fact that only
the second derivative of the wave function at origin depends on $n$
\begin{eqnarray}\label{psi_n}
 \Psi_{ns}(r\to 0) &\simeq&
 \frac{(Z\alpha m)^{3/2}}{\pi^{1/2}n^{3/2}}\times
 \Biggr\{1-(Z\alpha m r)+\frac{(Z\alpha m r)^2}{2}\nonumber\\
 &~&\qquad+\frac{1-n^2}{n^2}\,\frac{(Z\alpha m r)^2}{6}+\dots
 \Biggr\}
\end{eqnarray}
and the $n$-dependent coefficient is relatively small.
Under these circumstances we consider the partial results in Eqs.~(\ref{numpart}) as a confirmation
of direct calculation of the self-energy \cite{yero2001} and vacuum polarization (see Eq.~(\ref{numcvp})).

A summary of the contributions of the fourth order terms is presented in Table~\ref{17Tab14}.


\begin{table}[bth]
\caption{Fourth order QED contributions to the $D_{21}$ in hydrogen,
deuterium and helium-3 ion.}
\label{17Tab14}
\begin{center}
\def\arraystretch{1.4}
\setlength\tabcolsep{5pt}
\begin{tabular}{lccc}
\hline
Contribution                        &  H~~      & D~~        & $^3$He$^+$~~  \\
&[kHz]~~&[kHz]~~&[kHz]~~\\
\hline
$(Z\alpha)^4 E_F $                 & ~0.005\,6~~~~~ &  ~0.001\,3~~~~~&-0.543~~~~~ \\
$\alpha^2(Z\alpha)^2 E_F $         & ~0.003\,3(16)  &  ~0.000\,8(4) & -0.069(35) \\
$\alpha(Z\alpha)^2{m\over M} E_F $ & -0.003\,1(15)  &  -0.000\,4(2) & ~0.022(11) \\
$\alpha(Z\alpha)^3 E_F $ (SE)      & ~0.008\,3(10)  &  ~0.001\,9(2) & -0.395(37) \\
$\alpha(Z\alpha)^3 E_F $ (VP)      & ~0.003\,0~~~~~ &  ~0.000\,7~~~~& -0.145~~~~~ \\
$(Z\alpha)^3{m\over M} E_F $       & ~0.000\,5(5)~  &  ~0.000\,1~~~~& -0.007(10) \\
\hline
Total: 4th order & ~0.0178(25) & ~0.0043(5)~ & -1.137(53) \\
\hline
\end{tabular}
\end{center}
\end{table}

\section{Nuclear-structure corrections to $D_{21}$ \label{SectNucl}}

The leading nuclear-structure corrections to $E^{\rm Nucl}_{\rm hfs}(1s)$ and
$E^{\rm Nucl}_{\rm hfs}(2s)$, being proportional to
the wave function at origin (see Eq.~(\ref{NuclPsi2})), cancel each other when
calculating the difference $D_{21}$. However, some higher-order nuclear effects
can shift $D_{21}$ and, in fact, they do.
The corrections related to the nuclear structure effects can be splitted
into three terms \cite{preliminary,17kars97}
\begin{equation}\label{completeNucl}
 D_{21}({\rm Nucl})=D_{21}^A+D_{21}^B+D_{21}^C\;,
\end{equation}
where
\begin{eqnarray}\label{d21nuc}
 D_{21}^A&=&\left(\ln2+{3\over16}\right)\cdot(Z\alpha)^2\cdot E^{\rm Nucl}_{\rm hfs}(1s)
 \label{ad21}\;,\\
 D_{21}^B&=&\left({7\over4}-{4\over3}\ln2\right)\cdot(Z\alpha)^2(mR_E)^2E_F
 \label{bd21}\;,\\
 D_{21}^C&=&-{{\zeta}\over 4}\cdot(Z\alpha)^2(mR_E)^2E_F\label{cd21}\;.\\
\end{eqnarray}
Here
\begin{equation}
 \zeta =\left(\frac{R_M}{R_E}\right)^2 - 1
\end{equation}
is a ratio of quadratic magnetic and electric charge radii.

Let us discuss origin and accuracy of the nuclear-struc\-ture corrections.
To find the first term ($D^A_{21}$) one has to somehow determine a value of
the nuclear contribution to the ground state {\em hfs} separation,
$E^{\rm Nucl}_{\rm hfs}(1s)$, which contains three kinds of terms:
\begin{itemize}
 \item nuclear-finite-size effects of order $(Z\alpha)^3(mR)E_F$, where
 $R\sim R_E\sim R_M$;
 \item nuclear polarizability corrections;
 \item nuclear recoil corrections of order $(Z\alpha)^3(m/M)\ln(mR)$.
\end{itemize}

The correction for the $1s$ state was studied for hydrogen and deuterium.
In the case of the hydrogen the first term is dominant and cannot be calculated
with accuracy better than 20\% because of lack of knowledge of the proton magnetic
form factor at low momentum transfer \cite{17kars99,17kars97}. The proton polarizability
cannot be successfully estimated and delivers an essential contribution to the value of
$E^{\rm Nucl}_{\rm hfs}(1s)$ in hydrogen. We expect that a theoretical uncertainty of
nuclear contribution to the $1s$ {\em hfs} is at least 20\% of its value.

Deutron is a weekly bound nucleus and
the dominant nuclear effect for the hyperfine separation in the ground state of deuterium
is related to the deutron polarizability.
The nuclear correction was estimated in Ref.~\cite{deut_nucl} as 43 kHz,
however, the uncertainty is not presented there. We expect that the uncertainty lies between 10 and 30 kHz.
Our assumption is based on examination of the logarithmic approximation used in Ref.~\cite{deut_nucl}.
Let us concentrate our analysis on two corrections of $-19$ and $+11$~kHz which were found in
the logarithmic approximation. They are proportional to $\ln(m_p/\kappa)$, where $\kappa\simeq 45.7$ MeV is
the inverse deutron size
and $m_p$ is the proton mass. The validity of the logarithmic approximation suggests that the logarithm
is big enough, but that is not really a case: $\ln(m_p/\kappa)\simeq 3.0$.
We expect that the uncertainty of such an approximation lies between 10 and 30 kHz,
that depends on possible correlations between these two logarithmic contributions.

To the best of our knowledge there are no results published on the nuclear contributions
to the hyperfine separation in the tritium atom and the helium-3 ion. Due to lack of accurate calculations
for $E^{\rm Nucl}_{\rm hfs}(1s)$ we estimate the nuclear structure contribution to the $1s$ {\em hfs}
interval
in all atoms discussed above by a comparison of experimental data with a result of the QED calculations
(see Table~\ref{T1shfs} and Eqs.~(\ref{NuclPsi1}, \ref{NuclPsi2}, \ref{Nucl1s1}, \ref{Nucl1s2}).

All these corrections to $E^{\rm Nucl}_{\rm hfs}(1s)$
are related in the leading order to the two-photon
exchange with a hard-momentum exchange loop ($k\gg Z\alpha m$).
Their calculation is similar
to that for $\alpha(Z\alpha)^3E_F$ terms. Some of
these two-photon contributions can induce additional terms with
derivatives but that will involve an additional suppressing factor $m/k$. The
factor $m/k$ is small for the finite-size and polarizability and for a
logarithmic part (with $\ln(m R)$) of the nuclear recoil contribution. It is about unity only for
a part of the nuclear recoil contribution related to low momentum transfer $k\sim m$ and $k<m$,
however, we have calculated those corrections of order $(Z\alpha)^3(m/M)$ in the leading
logarithmic approximation in Eq.~(\ref{leadrec}). We expect these contributions
are relatively small, because the recoil effects are not dominant in
the $D^A_{21}$ term and because of small numerical coefficient for state-dependent terms in
the hydrogenic wave function at origin (see Eq.~(\ref{psi_n})). The $\gamma$-matrix structure is close to that in the case
of the vacuum polarization where the terms with derivatives induce
numerically small contributions. We estimate the uncertainty of such an
approximation for $D^A_{21}$ as 10\%.

To verify the expression for $D^A_{21}$ we also compare value of $E^{\rm Nucl}_{\rm hfs}(1s)$
with that for $E^{\rm Nucl}_{\rm hfs}(2s)/8$ (see Table~\ref{T2sNucl}). The latter was found via
a comparison of a pure QED theoretical expression (cf. Eqs. (\ref{QED1sF},\ref{QED1s},\ref{d21mod})
\begin{eqnarray}\label{QED2s}
E_{\rm hfs}^{\rm QED}(2s) &=& \frac{E_F}{8} \bigl(1+Q_{QED}(2s)\bigr)\;,\nonumber \\
Q_{\rm QED}(2s)&=&a_e+\left\{\frac{17}{8}(Z\alpha)^2+ \alpha(Z\alpha)\left(\ln2-\frac{5}{2}\right)\right.\nonumber\\
&+&{\alpha (Z \alpha )^2\over \pi}\left[-\frac{2}{3}\ln{1\over(Z\alpha)^2}
\left(\ln{\frac{1}{(Z\alpha)^2}}\right.\right.\nonumber \\
&+&\left.8\ln2-\frac{1541}{240}\right) + 11.901\,106\ldots \nonumber \\
&-&\left.\left.-\frac{247}{450}\right]+0.7718(4)\,\frac{\alpha^2(Z\alpha)}{\pi}\right\}\,.
\end{eqnarray}
withexperimental data. The results for the nuclear contributions to
the $1s$ state in Table~\ref{T1shfs} and for the
$2s$state in Table~\ref{T2sNucl} agree with each other.


\begin{table}[bth]
\caption{$2s$ hyperfine splitting in light atoms. $\Delta E_{\rm hfs}=E_{\rm hfs}^{\rm exp}-E_{\rm hfs}^{\rm QED})$
and measured in ppm in respect to $E_F/8$.
}
\label{T2sNucl}
\begin{center}
\def\arraystretch{1.4}
\setlength\tabcolsep{5pt}
\begin{tabular}{lccc}
\hline
Atom, & $E^{\rm exp}_{\rm hfs}$& $E^{\rm QED}_{\rm hfs}$& $\Delta E_{\rm hfs}$ \\
state&[kHz]&[kHz]&[ppm]\\
\hline
H, $2s$ & ~~177\,556.785(29), \protect\cite{17rothery}& ~~177\,562.7~~ & -33\\
H, $2s$ & ~~177\,556.860(50), \protect\cite{17h2s}& &-32\\
D, $2s$ & ~~~40\,924.439(20), \protect\cite{17d2s}& ~~~40\,918.81 &137\\
$^3$He$^+$, $2s$ & -1\,083\,354.981(9), \protect\cite{17prior}& -1\,083\,594.7~~& 221\\
$^3$He$^+$, $2s$ & -1\,083\,354.99(20), \protect\cite{17he2s}& &221\\
\hline
\end{tabular}
\end{center}
\end{table}

Two other nuclear contributions, $D_{21}^B$ and $D_{21}^C$ (see Eq.~(\ref{bd21}) and Eq.~(\ref{cd21})), are smaller
than $D_{21}^A$ and their evaluation is similar to
that for the $\alpha(Z\alpha)^2E_F$ contributions and completely understood.
They were derived in Refs.\cite{preliminary,17kars97}
with help of some effective potentials, and the result does not depend on any nuclear models.
The result for the nuclear contribution in Eq.~(\ref{completeNucl}) can be presented in a form
slightly different from Eqs.~(\ref{completeNucl}, \ref{ad21}, \ref{bd21}, \ref{cd21})
\begin{eqnarray}\label{d21nuc1}
D_{21}({\rm Nucl})&=&\left(\ln2+{3\over16}\right)
\cdot(Z\alpha)^2\cdot \Delta E^{\rm Nucl}_{\rm hfs}(1s)\nonumber\\
&+&\left({21\over8}-2\ln2\right)
\cdot\frac{\Delta E_{\rm Lamb}^{\rm Nucl}(1s)}{(Z\alpha)^2m}\,E_F\;,\nonumber\\
&-&{3\over 8}\,\zeta
\cdot\frac{\Delta E_{\rm Lamb}^{\rm Nucl}(1s)}{(Z\alpha)^2m}\,E_F\;,\nonumber
\end{eqnarray}
which is more useful for phenomenological applications. We check this expression for the effects caused
by a distribution of the nuclear charge and magnetic moment within some models in Appendix~\ref{appendixModel}
and confirm it. The results on nuclear contributions to the $D_{21}$ in the light hydrogen-like atoms
are presented in Table~\ref{17Tab1n}.

%
\begin{table}[bth]
\caption{Nuclear-structure contributions to the $D_{21}$ in hydrogen, deuterium and helium-3 ion.
}
\label{17Tab1n}
\begin{center}
\def\arraystretch{1.4}
\setlength\tabcolsep{5pt}
\begin{tabular}{lccc}
\hline
Contribution                        &  H      & D        & $^3$He$^+$  \\
& [kHz]& [kHz]& [kHz]\\
\hline
$D_{21}^A$ & $-0.002\,2(2)$ & 0.002\,1(2) & 0.360(36)  \\
$D_{21}^B$ & 0.000\,3 &  0.000\,4 & $-0.028\,5$ \\
$D_{21}^C$ & $-1\cdot10^{-4}\cdot\zeta$ & $-1.3\cdot10^{-4}\cdot\zeta$ & $8.6\cdot10^{-3}\cdot\zeta$\\
\hline
$D_{21}({\rm Nucl})$ & $-0.002$ & 0.002\,6(2) & 0.332(36) \\
& $-1\cdot10^{-4}\cdot\zeta$ & $-1\cdot10^{-4}\cdot\zeta$ & $+9\cdot10^{-3}\cdot\zeta$\\
\hline
\end{tabular}
\end{center}
\end{table}

The final theoretical results
\begin{equation}
D_{21}({\rm theor}) = D_{21}({\rm QED})+D_{21}({\rm Nucl})
\end{equation}
for hydrogen, deuterium and helium-3 ion
are summarized in Table~\ref{Td21new}. The table
contains also the experimental results. The main sources of uncertainty of
the theoretical calculations are related to
\begin{itemize}
\item the use of logarithmic
approximation in evaluation of higher order QED corrections of order
$\alpha^2(Z\alpha)^3E_F$, $\alpha(Z\alpha)^2(m/M)E_F$ and $(Z\alpha)^3(m/M)E_F$;
\item calculation of higher-order nuclear effects.
\end{itemize}
The recoil contributions in order $\alpha(Z\alpha)^2(m/M)E_F$ and $(Z\alpha)^3(m/M)E_F$
also limit accuracy of the calculations of the hyperfine splitting in the
ground state of muonium and positronium and we overview theory of these
two quantities in the next section.

table 6
\begin{table*}[bt]
\caption{Value of $D_{21}$ in hydrogen, deuterium and helium-3 ion.
Results for nuclear correction and theory are for $\zeta=0$.
$\Delta({\rm exp-th})=D_{21}({\rm exp})-D_{21}({\rm theor})$ and $\sigma$ is a
final uncertainty of $\Delta({\rm exp-th})$. The final nuclear contribution $D_{21}({\rm Nucl})$
here is presented at $\zeta=0$}
\label{Td21new}
\begin{center}
\def\arraystretch{1.4}
\setlength\tabcolsep{5pt}
\begin{tabular}{lccc}
\hline Value         &  H~~~~~~   & D~~~~~~  & $^3$He$^+$~~~~~~ \\ 
\hline 
$D_{21}({\rm exp})$~[kHz]&48.53(23),~\protect\cite{17rothery} 
&11.16(16),~\protect\cite{17d2s}& -1\,189.979(71),~\protect\cite{17prior}\\
$D_{21}({\rm QED})$~[kHz]&48.955(3)~~~~~~&11.309\,9(5)~~~~&-1\,190.400(53)~~~~~~\\
$D_{21}({\rm Nucl})$ [kHz] & -0.002~~~~~~~~~~ & 0.002\,6(2)~~~   & ~0.332(36) \\
$D_{21}({\rm theor})$ [kHz] & 48.953(3)~~~~~ &  11.312\,5(5)~~~ & -1\,190.067(64)~~~~~\\ 
$\Delta({\rm exp-th})$  [kHz] & -0.42(23)~~~~~~ &  -0.15(16)~~~~~ & ~0.09(10)  \\
$\Delta({\rm exp-th})/\sigma$ & -1.8~~~~~~ &  -1.0~~~~~~ & 0.9~~~~~~~ \\ 
$\sigma/E_F$ [ppm]& 0.16~~~~~~ &  0.49~~~~~~ & 0.01~~~~~~ \\ \hline
\end{tabular}
\end{center}
\end{table*}

\section{Hyperfine structure in pure leptonic atoms}

The theoretical expression for the {\em hfs} interval in the muonium ground state can be presented
in the form
\begin{eqnarray}
E_{\rm hfs}({\rm theor})&=&E_F\left(1+Q\right)\nonumber\\
&=&E_F\left(1+a_e+Q_2 +Q_3+Q_4+Q_h+Q_w\right)\;,\nonumber
\end{eqnarray}
and the results for the QED contributions of the second ($Q_2$), third
($Q_3$) and fourth order ($Q_4$), for the hadronic ($Q_h$) and weak
contributions ($Q_w$) are reviewed in Ref. \cite{CEK} (see also
Refs.~\cite{icap,17ZP,EGS}) and we follow consideration there.
The hadronic contribution is taken from Ref.~\cite{CEK,KShad}.
The results are collected in Table~\ref{17TabMu}.


%
\begin{table}[bth]
\caption{Muonium hyperfine structure}
\label{17TabMu}
\begin{center}
\def\arraystretch{1.4}
\setlength\tabcolsep{5pt}
\begin{tabular}{lcc}
\hline
Term  &  Fractional & $\Delta E$ \\
   &  contribution &  [kHz] \\
\hline
Fermi energy & ~~1.000\,000\,000~~~~~ &4.459\,031.92(51)~~~~~~\\
$a_e$ & ~~0.001\,159\,652~~~~~~&5\,170.93~~~~~~\\
2nd order QED & - 0.000\,195\,815~~~~~ & - 873.15~~~~~~\\
3rd order QED & - 0.000\,005\,923~~~~~ & - 26.41~~~~~\\
4th order QED & - 0.000\,000\,123(49)~& ~~- 0.55(22)\\
Hadronic effects & ~~0.000\,000\,054(1)~~ & 0.24~~\\
Weak int. &- 0.000\,000\,015~~~~~~ &0.06~~\\
\hline
Total & ~~1.000\,957\,830(49) &4\,463\,302.91(51)(22)\\
\hline
\end{tabular}
\end{center}
\end{table}

The dominant QED contribution to the uncertainty (0.22~kHz) comes from unknown
non-leading terms in orders $\alpha(Z\alpha)^2m/M$ and $(Z\alpha)^3m/M$,
which are estimated by a half-value of the leading double logarithmic
corrections \cite{17ZP,17Zh}, despite some terms beyond the double
logarithms are known (see Ref.~\cite{CEK} for discussion).

To find an absolute value one has to determine the Fermi energy in Eq.~(\ref{fermiE}) which
contains the magnetic moment of the muon. The most accurate value of it
can be obtained from study of the ground state hyperfine structure in the
magnetic field \cite{MuExp}. The related uncertainty is 0.51~kHz.
Opposite to the theory of $D_{21}$, the Fermi energy has to be calculated very
accurately and its value depends on our choice of a value of the fine structure
constant. Here we use an original result from study of anomalous magnetic
moment of the electron $\alpha^{-1}=137.035\,999\,58(52)$
\cite{kinoshita}. The related uncertainty is only 0.03~kHz, however, scattering
of various results for the fine structure constants (see e.g. Ref.~\cite{17mohr})
corresponds to a much bigger uncertainty.

The positronium {\em hfs} interval can be calculated and measured less
accurately than that in muonium. However, it provides us with a sensitive test of the same
recoil corrections as in the case of muonium and $D_{21}$. The recoil effects in positronium
are essentially bigger than in other atoms, because
$M=m$, and they may be studied in detail. The positronium hyperfine splitting in the ground state can be
presented in the form
\begin{eqnarray}
 E_{\rm hfs}({\rm theor})&=&E_F\left(1+q_1\,\alpha+q_2\,\alpha^2+q_3\,\alpha^3\right)
 \nonumber\\
 &=&E_F\left(1+Q_1+Q_2 +Q_3\right)\;,\nonumber
\end{eqnarray}
where coefficients slightly depend on $\alpha$ containing
$\ln\alpha$. The Fermi energy is defined in positronium in a different way
(comparing with Eq.~(\ref{fermiE}) for hydrogen and muonium)
\begin{equation}
 E_F({\rm Ps})/h={7\over 6}\,\alpha^2\,c\,Ry
\end{equation}
because of annihilation effects and a symmetric treatment of
magnetic moments of the electron and the nucleus (posi\-tron). The results
are summarized in Table~\ref{17TabPs} (see \cite{Adk,Hoa,PK,Czar,EGS,a7ps}
and references there). The third order corrections appear to be large because of
a double logarithmic enhancement \cite{17Zh} ($\ln^2\alpha\simeq 24$). The value of $q_3$ is
calculated with taking into account recent results on $\alpha^7m\ln\alpha$
correction \cite{a7ps}, however, the uncertainty is estimated by a
half-value of the leading $\alpha^7m\ln^2\alpha$ \cite{17Zh} (see Ref.~\cite{CEK} for
discussion).


%
\begin{table}[bth]
\caption{Positronium hyperfine splitting}
\label{17TabPs}
\begin{center}
\def\arraystretch{1.4}
\setlength\tabcolsep{5pt}
\begin{tabular}{lccc}
\hline
Term  &  $q$ & $Q$ & $\Delta E$ [MHz] \\
\hline
Fermi energy & 1 & 1.000\,000\,0~~~~~& 204\,386.6~~~~\\
1st order QED &-0.674\,16 & - 0.004\,919\,6~~~~~~~&-1\,005.5\\
2nd order QED &1.084& 0.000\,057\,7~~~~~&~~~~11.8\\
3rd order QED &-15.6&-0.000\,006\,1(22) &~~~~~~~-1.2(5)\\
\hline
Total &&~0.995\,132\,1(22) & 203\,391.7(5)\\
\hline
\end{tabular}
\end{center}
\end{table}

\section{Summary}

The results for the precision calculations of the hyperfine structure in the light
hydrogen-like atoms are summarized in Table~\ref{TSummary}. One can see that investigations of
the difference $D_{21}$ provide very accurate tests of the bound state QED
calculations. We consider study of $D_{21}$ as a test of a state-dependent sector of theory
of the hyperfine splitting of the $1s$ and $2s$ states and so the fractional accuracy of
such theory is related to the $E_F$, the leading contribution to the $1s$ {\em hfs}.
The accuracy of comparison of theory and experiment can be characterized by a standard
deviation $\sigma$ which contains contributions to uncertainty from both: theory and experiment.
The final uncertainty is found to be for $D_{21}$ as small as few part of $10^7$ in the case of hydrogen
and deuterium and even better in the
case of helium ion: a part of $10^8$. That is competitive with
other tests of the bound state QED and in order to clarify advantages
and disadvantages of studying $D_{21}$ let us list main problems which theoretical
calculations have met by now:
\begin{itemize}
 \item there are two essential problems of the bound state QED:
 \begin{itemize}
  \item evaluation of higher-order recoil corrections (that is mainly a problem of all
  QED calculations for the hyperfine structure including $D_{21}$);
  \item evaluation of higher-order two-loop corrections (that is rather a problem of the Lamb shift
  calculation and only one value related to the hyperfine structure, $D_{21}$, is sensitive to such corrections);
 \end{itemize}
 \item there are two other problems related to other part of physics:
 \begin{itemize}
  \item determination of the fundamental constants (like e.g. determination of the fine structure constant
  and magnetic moment of muon needed to calculate the Fermi energy $E_F$);
  \item nuclear structure, which affect energy levels and, in particular, shifts
values of the Lamb shift and the hyperfine separation.
 \end{itemize}
\end{itemize}


\begin{table*}[bt]
\caption{Hyperfine splitting: precision tests of the bound state QED.
The final uncertainty $\sigma$ includes
contributions from both: theory and experiment. References for the
$D_{21}$ are presented for the both states: $2s$ and $1s$.}
\label{TSummary}
\begin{center}
\def\arraystretch{1.4}
\setlength\tabcolsep{5pt}
\begin{tabular}{lcccccc}
\hline
Atom  &  Value & Exp.& Theor & $\Delta$ & $\Delta/\sigma$ & $\sigma/E_F$ \\
 & & [kHz] & [kHz] & [kHz] &   & [ppm] \\
\hline
Mu & $E_{\rm hfs}(1s)$ & 4\,463\,302.78(5), \protect\cite{MuExp} & 4\,463\,302.91(56)& & & 0.12 \\
Ps & $E_{\rm hfs}(1s)$ & $203\,389.1(7)\cdot 10^3$, \protect\cite{PsExp1} & $203\,391.9(5)\cdot 10^3$& -2.8(9)& -3.3 & 3.4 \\
Ps & $E_{\rm hfs}(1s)$ & $203\,387.5(16)\cdot 10^3$, \protect\cite{PsExp2} & & -4.4(17) & -2.6 & 7.9 \\
 \hline
H  & $D_{21}$ & 48.53(23), \protect{\cite{17rothery}}/\protect{\cite{cjp2000}} & 48.953(3) &-0.42(23) & -1.8 & 0.16 \\
H  & $D_{21}$ & 49.13(40), \protect{\cite{17h2s}}/\protect{\cite{cjp2000}} & & 0.18(40)& 0.4 &  0.28 \\
D  & $D_{21}$ & 11.16(16), \protect{\cite{17d2s}}/\protect{\cite{17d1s}} & 11.312\,5(5)&-0.15(16) & -1.0 & 0.49 \\
$^3{\rm He}^+$  & $D_{21}$ & -1\,189.979(71), \protect{\cite{17prior}}/\protect{\cite{17he1s}} & -1\,190.068(64)& 0.09(10)& 0.9& 0.01 \\
$^3{\rm He}^+$  & $D_{21}$ & -1\,190.1(16), \protect{\cite{17he2s}}/\protect{\cite{17he1s}} & & 0.03(160) &  -0.02 & 0.18 \\
\hline
\end{tabular}
\end{center}
\end{table*}

The difference $D_{21}$ happens to be an only value that is sensitive to
both higher-order corrections: recoil and two-loop and that is not
sensitive to problems beyond QED (determination of the
fundamental constants and nuclear structure).
Tests of QED are sometimes considered as a search for new physics beyond the Standard model.
Such exotic contributions are rather expected to be proportional to $1/n^3$ and
must vanish for $D_{21}$ in the leading order. The next-to-leading terms could contribute but they
effectively would be taken into account in $D_{21}^{\rm Nucl}$ being included into $E_{\rm hfs}^{\rm Nucl}(1s)$.
That fact makes the difference $D_{21}$ useful for a very specific test of the bound state QED, a test which
involves no problem beyond QED.

Our theoretical predictions appear to be in a fair agreement with four of five accurate measurements (see Table~\ref{TSummary}), while
a minor discrepancy of $1.8\;\sigma$ with the most recent result from Ref.~\cite{17rothery} is observed. Because 
of agreement with other data and espetially with the most accurate result for helium ion \cite{17prior} we expect 
that the problem of this minor discrepancy comes from the experimental side. One can see
that there have been no improvement in microwave measurements of the $2s$ {\em hfs} for the last few decades. We expect
that some progress is still possible and that it is now also possible to perform an optical measurement of this quantity via
comparison of different $1s-2s$ transitions in hydrogen and deuterium, some of which were measured recently very precisely
\cite{1s2s}.

\section*{Acknowledgments}

The authors would like to thank Andrzej Czarnecki, Simon Eidelman, Eric
Hessels, Dan Kleppner, Mike Prior and Valery Shelyuto for stimulating
discussions. An early part of this work was done during a short but fruitful visit of SGK to
University of Notre Dame and he is very grateful to Jonathan Sapirstein for
his hospitality, stimulating discussions and participation in the early
stage of this project. The work was supported in part by RFBR grant
00-02-16718, NATO grant CRG 960003, DAAD and by Russian State Program
``Fundamental Metrology''.

\appendix

\section{The vacuum polarization contribution to the $2s$ hyperfine splitting \label{appendixVP}}

An exact relativistic expression for the vacuum polarization correction to
{\em hfs} in the ground state of a hydrogen-like atom with a point-like nucleus
was derived in Ref. \cite{jetp00,17kis}. Using
the same method we can calculate the vacuum polarization contribution for the $2s$-state.
As in the case of the $1s$ state \cite{jetp00}, we study a more general case
considering an orbiting particle with the mass $m$ different from the electron mass $m_e$ which is
related here to a particle in the vacuum loop (see. Fig.~1). That
offers an opportunity to perform some additional tests of our results.

The diagrams contributing to the {\em hfs} separation are presented in Fig.~1. We obtain
\begin{eqnarray*}
&&\Delta E_2(\mbox{VP--TU})=\frac{\alpha}{\pi}\,E_T(2s)\nonumber\\
&\times&\frac{1}{2-5\eps+2\eps^2-2\En(1+\En)(5-2\eps)}\nonumber\\
&\times&\biggl\{-4\En(1+\En)(3-2\eps)^2(1-\eps)\,J_{10}(\kp_2)\nonumber\\
&-&(1+6\En+6\En^2)(3-2\eps)^2(1-2\eps)\,J_{20}(\kp_2)\nonumber\\
&+&2(1+2\En)^2 (1-\eps)(5-13\eps+6\eps^2) \, J_{30}(\kp_2)\nonumber\\
&-&(1+2\En)^2 (1-\eps) (3-8\eps+4\eps^2) \, J_{40}(\kp_2)\biggr\}\nonumber\\
\end{eqnarray*}
for the single-potential contribution related to Fig.~1a and
\begin{eqnarray*}
&&\Delta E_2(\mbox{VP--U$\cdot$T}) =
-\frac{\alpha}{\pi} \, E_T(2s)\\
&\times&\frac{\En\Zab^2(3-2\eps)^2}
{\eps(1-2\eps)^2(5+4\En-2\eps){(2-\eps)^2}}\\
&\times&\biggl\{-\frac{2(1+\eps)(2+2\En-5\eps+2\eps^2)}{1-2\eps} \, J_{10}(\kpt_2)\\
&+&\frac{4-9\eps^2+4\eps^3}{1-\eps} \, J_{20}(\kpt_2)\\
&&~~+\frac{4-21\eps+23\eps^2-8\eps^3}{(1-\eps)^2} \,\En\, J_{20}(\kpt_2)\\
&-&\frac{2\bigl(-18+31\eps-19\eps^2+4\eps^3\bigr)}{3-2\eps} \, J_{30}(\kpt_2)\\
&&~~-\frac{2\bigl(-6+9\eps-10\eps^2+4\eps^3\bigr)}{3-2\eps} \,\En\, J_{30}(\kpt_2)\\
&-&2\Bigl[10-9\eps+2\eps^2+\En(13-14\eps+4\eps^2)\Bigr]\, J_{40}(\kpt_2)\\
&+&2(1+2\En){{(2-\eps)}^2} \, J_{50}(\kpt_2)\\
&+&\frac{4(2-\eps)(1+\En-\eps)}{1-\eps} \, J_{21}(\kpt_2)\\
&-&8(1+\En)(2-\eps) \, J_{31}(\kpt_2)\\
&+&4(1+2\En)(2-\eps) \, J_{41}(\kpt_2)\biggr\}
\end{eqnarray*}
for the double-potential term in Fig.~1b.
Here we mainly follow notations of Refs.~\cite{karsVP,jetp00} and, in particular, we introduce
the relativistic Fermi-Breit energy \cite{17brei}
\begin{eqnarray*}
E_T(2s)&=&E_F \times {\epsilon\over2(Z\alpha)^2}\\
&~&\quad\times\frac{\Bigl[(1+\En)(5-2\epsilon)-1\Bigr]}
{(1+2\En)(1-\epsilon)(2-\epsilon)(3-8\epsilon+4\epsilon^2)}\\
&\approx& \frac{E_F}{8} \left( 1 + \frac{17}{8}\Zab^2 + \dots \right)\;,
\end{eqnarray*}
where
\begin{eqnarray*}
\eps&=&1-\sqrt{1-\Zab^2)}={\Zab^2\over 2} \left(1+{\Zab^2\over 4}+\ldots\right)\,,\\
\En&=&\sqrt{\frac{2-\eps}{2}}\simeq 1-{\Zab^2\over 8}+\ldots\,,\\
\kp&=&\frac{\Za\,m}{m_e}\,,\\
\kpt_2&=&\frac{\kp}{2\En}={\kp\over 2}\left(1+{\Zab^2\over 8}+\ldots\right)\,.
\end{eqnarray*}
Basic integrals $J_{mn}$ are defined as
\begin{eqnarray*}
J_{mn}(\kappa) =
\int_0^1dv&\;&\frac{v^2(1-v^2/3)}{1-v^2}
\left( \frac{\kappa \sqrt{1-v^2}}{1+\kappa \sqrt{1-v^2}} \right )^{m-2\epsilon}\\
&\times&\ln^n \left( \frac{\kappa \sqrt{1-v^2}}{1+\kappa \sqrt{1-v^2}} \right)\;.
\end{eqnarray*}
They can be expressed in terms of the beta-function and the hypergeometric function as follows \cite{karsVP,jetp00}
\begin{eqnarray*}
&&J_{m0}=\frac{1}{2}\kp^{m}\,B\big(3/2,m/2\big)\\
&\times&{_3F_2}\big(m/2,\,m/2+1/2,\,m/2;\;1/2,\,m/2+3/2;\;\kp^2\big)\\
&-&\frac{m}{2}\,\kp^{m+1}\,B\big(3/2, m/2+1/2\big)\\
&\times&{_3F_2}\big(m/2+1, m/2+1/2, m/2+1/2;\,3/2, m/2+2;\, \kp^2\big)\\
&-&\frac{1}{6}\kp^{m}\,B\big(5/2,m/2\big)\\
&\times&{_3F_2}\big(m/2,\, m/2+1/2,\, m/2 ;\;1/2,\, m/2+5/2 ;\; \kp^2\big)\\
&+&\frac{m}{6}\,\kp^{m+1}\,B\big(5/2, m/2+1/2\big)\\
&\times&{_3F_2}\big(m/2+1, m/2+1/2, m/2+1/2;\,3/2, m/2+3;\, \kp^2\big)
\end{eqnarray*}
and
\[
J_{mn}=\frac{\partial^n J_{m0}}{\partial m^n}\;.
\]

In the case of an electronic atom ($m=m_e$) and small $\Za$ we arrive to
a result in Eq.~(\ref{eFig1}), which reproduces all known
terms of the expansion over $\Za$ \cite{VPall,17zwanziger,17ZP} and presents a new
contribution in order $\alpha(Z\alpha)^3E_F$.

One can also study the vacuum-polarization contribution to the {\em hfs} in
muonic atoms putting $m=m_\mu$. In the case of low $Z$ and
arbitrary $\kappa=\Za m_\mu/m_e\simeq 1.5Z$ we reproduce the non-relativistic limits \cite{ejp98}.
For the case of small $\Za$ and large $\kp$ the result can be
presented as an expansion over $\Za$ and $\kp^{-1}$:
\begin{eqnarray*}
 \Delta E_2(\mbox{VP})&=&\frac{\alpha}{\pi} \frac{E_F}{8}\left\{\left[\frac{8}{3}\ln(\kp)
 +\frac{4\pi^2}{9}-\frac{85}{18}+\frac{49}{\kp^2}\right]\right.\nonumber\\
 &+&\Zab^2\left[\frac{17}{2}\ln(\kp)+\frac{3\pi^2}{2}-\frac{37}{3}\right.\nonumber\\
&&\qquad\qquad-\frac{8}{3}\psi^{\prime\prime}(2)+\left.\left.\frac{303}{4\kp^2}\right]+\dots\right\}\;,\label{medium}
\end{eqnarray*}
where $\psi(z)$ is the logarithmic derivative of the $\Gamma$-function.
The logarithmic part of the correction can be easily found within the
effective charge approach (cf. \cite{karsVP,jetp00}), results agree with
(\ref{medium}) and that is an additional confirmation of our result.

\section{A model-dependent calculation of the finite-nuclear-size corrections to energy levels \label{appendixModel}}

Here we study the contributions to $E_{\rm Lamb}^{\rm Nucl}(1s)$, $E_{\rm hfs}^{\rm Nucl}(1s)$ and
$D_{21}$ related to the distribution of the nuclear charge and the nuclear magnetic moment. The
distribution is described by the nuclear electric and magnetic form factors
\begin{eqnarray*}
 G^a_E(q^2) &=& 1 +\beta_E\left[
 \left(\frac{\Lambda_E^2}{q^2+\Lambda_E^2}\right)^a-1\right]\;,\nonumber\\
 G^a_M(q^2) &=& \mu\left\{1 +\beta_M
 \left[\left(\frac{\Lambda_M^2}{q^2+\Lambda_M^2}\right)^a-1\right]\right\}\;.
\end{eqnarray*}
Parameters $\Lambda$ and $\beta$ are free parameters, however, we consider here only linear in $\beta$ contributions.
Since we intend to verify a model independent expression in Eq.~(\ref{d21nuc1}),
it is not important that the distribution above is not quite a real one.

We perform calculations with the Dirac wave functions and expand the results over
$(Z\alpha)$ and $m/\Lambda$. In the case of $a=1$ the nuclear corrections are
\begin{eqnarray*}
E_{\rm Lamb}^{\rm Nucl}(1s)&=&4(Z\alpha)^4  \beta_E \left( \frac{m}{\Lambda_E} \right)^2m\;,\nonumber\\
E_{\rm hfs}^{\rm Nucl}(1s)&=&-4Z\alpha \left(\frac{\beta_E m}{\Lambda_E}+\frac{\beta_M m}{\Lambda_M}\right)E_F\;,\nonumber\\
D_{21}({\rm Nucl})&=&-\left(\frac34 + 4\ln 2\right)(Z\alpha)^3\left(\frac{\beta_E m}{\Lambda_E}
+ \frac{\beta_E m}{\Lambda_M}\right)E_F\nonumber\\
&&\qquad+( 12-8\ln 2)(Z\alpha)^2\beta_E  \left( \frac{m}{\Lambda_E} \right)^2
E_F\nonumber\\
&&\qquad\qquad-\frac32(Z\alpha)^2\beta_M \left(\frac{m}{\Lambda_M}\right)^2E_F\\
\end{eqnarray*}
and
\begin{eqnarray*}
R_E&=&{6\beta_E\over\Lambda_E^2}\;,\nonumber\\
R_M&=&{6\beta_M\over\Lambda_M^2}\;.
\end{eqnarray*}

We note that in the case of $\beta_E=\beta_M=1$,
$\Lambda_E=\Lambda_M$ we arrive at $a=2$ to the so-called dipole model commonly
used as an approximation for the proton internal structure. However, since we calculate the linear
in $\beta$ terms the well known result for the hydrogen {\em hfs} cannot be reproduced.
In the case of $a=2$ we find
\begin{eqnarray*}
E_{\rm Lamb}^{\rm Nucl}(1s)&=&8(Z\alpha)^4\beta_E\left(\frac{m}{\Lambda_E}\right)^2m\;,\nonumber\\
E_{\rm hfs}^{\rm Nucl}(1s)&=&-6Z\alpha\left(\frac{\beta_E m}{\Lambda_E} + \frac{\beta_M m}{\Lambda_M}\right)E_F\;,\nonumber\\
D_{21}({\rm Nucl})&=&-\left(\frac98 + 6\ln 2\right)(Z\alpha)^3 \left(\frac{\beta_E m}{\Lambda_E}
+ \frac{\beta_M m}{\Lambda_M}\right)E_F\nonumber\\
&&\qquad+( 24-16\ln 2 )(Z\alpha)^2\beta_E \left( \frac{m}{\Lambda_E} \right)^2 E_F\nonumber\\
&&\qquad\qquad-3(Z\alpha)^2\beta_M  \left( \frac{m}{\Lambda_M} \right)^2E_F
\end{eqnarray*}
and
\begin{eqnarray*}
R_E&=&{12\beta_E\over\Lambda_E^2}\;,\nonumber\\
R_M&=&{12\beta_M\over\Lambda_M^2}\;.
\end{eqnarray*}

The results for $a=1$ and $a=2$ confirm expression in Eq.~(\ref{d21nuc1}). Similar calculations can be performed
for any integer value of $a$. Let us also mention, that in principle any moments of a real distribution
($\langle R^n \rho_N(R)\rangle$)
can be reproduced by a finite sum of $G^a_E(q^2)$ and $G^a_M(q^2)$ over integer $a$ after
adjusting parameters $\Lambda(a)$ and $\beta(a)$.

\end{document}